\documentclass[12pt,preprint]{aastex}

\newcommand{\xrtposra}{\mbox{RA(J2000)=20$^{\rm h}$00$^{\rm m}$42$\fs7$}}
\newcommand{\xrtposdec}{\mbox{Dec(J2000)=$+09\degr$08$\arcmin$31\farcs4 }}

\newcommand{\ep}{E_{\rm peak}}

\begin{document}
\title{The {\it Swift} X-ray flaring afterglow of GRB 050607 }
\author{Claudio Pagani\altaffilmark{1,2}, David C. Morris\altaffilmark{2},
Shiho Kobayashi\altaffilmark{3}, Takanori Sakamoto\altaffilmark{4},
Abraham D. Falcone\altaffilmark{2}, 
Alberto Moretti\altaffilmark{1},
Kim Page\altaffilmark{5},
David N. Burrows\altaffilmark{2},
Dirk Grupe\altaffilmark{2}, 
Alon Retter\altaffilmark{2}, 
Judith Racusin\altaffilmark{2}, 
Jamie A. Kennea\altaffilmark{2}, 
Sergio Campana\altaffilmark{1}, Patrizia Romano\altaffilmark{1},
Gianpiero Tagliaferri\altaffilmark{1}, 
Joanne E. Hill\altaffilmark{4,6},
Lorella Angelini\altaffilmark{4},
Giancarlo Cusumano\altaffilmark{7},
Michael R. Goad\altaffilmark{5},
Scott Barthelmy\altaffilmark{3}, Guido Chincarini\altaffilmark{1},  
Alan Wells\altaffilmark{8}, Paolo Giommi\altaffilmark{9}, 
John A. Nousek\altaffilmark{2}, Neil Gehrels\altaffilmark{4} }

\altaffiltext{1}{INAF -- Osservatorio Astronomico di Brera, Via Bianchi 46, 23807 Merate, Italy; {\it pagani@astro.psu.edu}}
\altaffiltext{2}{Department of Astronomy \& Astrophysics, 525 Davey Lab., Pennsylvania State
University, University Park, PA 16802, USA}
\altaffiltext{3}{Astrophysics Research Institute, Liverpool John Moores University, Twelve Quays House, Birkenhead CH41 1LD, UK}
\altaffiltext{4}{NASA/Goddard Space Flight Center, Greenbelt, MD 20771, USA}
\altaffiltext{5}{Department of Physics and Astronomy, University of Leicester, Leicester LE1 7RH, UK}
\altaffiltext{6}{Universities Space Research Association, 10211 Wincopin Circle, Suite 500,
Columbia, MD, 21044-3432, USA}
\altaffiltext{7}{INAF -- Istituto di Astrofisica Spaziale e Fisica Cosmica Sezione di Palermo, Italy
Via U. La Malfa 153, 90146 Palermo, Italy}
\altaffiltext{8}{Space Research Centre, University of Leicester, Leicester LE1 7RH, UK}
\altaffiltext{9}{ASI Science Data Center, via Galileo Galilei, 00044 Frascati, Italy}

\begin{abstract}

The unique capability of the {\it Swift} satellite to perform a prompt and autonomous slew to a newly
detected Gamma-Ray Burst (GRB) has yielded the discovery of interesting new properties of
GRB X-ray afterglows, such as the steep early lightcurve decay  
and the frequent presence of flares detected up to a few hours after the GRB trigger.
We present observations of GRB 050607, 
the fourth case of a GRB discovered by {\it Swift} with
flares superimposed on the overall fading X-ray afterglow.
The flares of GRB 050607 were not symmetric as in previously reported cases,
showing a very steep rise and a shallower decay, similar
to the Fast Rise, Exponential Decay that are frequently observed
in the gamma-ray prompt emission.
The brighter flare had a flux increase by a factor of $\sim$~25,
peaking for 30~seconds at a count rate of approximately 30~counts~$s^{-1}$,
and it presented hints of addition short time scale activity 
during the decay phase.
There is evidence of spectral evolution during the flares.
In particular, at the onset of the flares the observed emission was harder,
with a gradual softening as each flare decayed.
The very short time scale and the spectral variability during the flaring 
activity are indicators of possible extended periods of energy 
emission by the GRB central engine. 
The flares were followed by a phase of shallow decay,
during which the forward shock was being 
refreshed by a long-lived central engine or 
by shells of lower Lorentz factors, 
and by a steepening after approximately 12~ks
to a decay slope considered typical of X-ray afterglows.  

\end{abstract}


\section{Introduction}

The {\it Swift} Explorer \citep{Gehrels}, designed
to discover and study Gamma-Ray Bursts (GRBs), was launched on
2004 November 20. 
When the Burst Alert Telescope (BAT, \citealt{Barthelmy}) triggers 
on a newly discovered GRB, the {\it Swift} satellite 
performs an autonomous slew to the GRB position determined on-board, 
allowing prompt follow-up 
observations with the two narrow-field instruments, 
the X-ray Telescope (XRT, \citealt{Burrows_xrt}) and 
the Ultraviolet/Optical Telescope 
(UVOT, \citealt{Roming}).

In the standard scenario the GRB emission is caused by
shocks in 
a relativistic expanding fireball \citep{model}.    
The internal shocks of the moving ejecta are responsible for the 
prompt gamma-ray emission,  while external shocks caused by the interaction of
the fireball with the surrounding medium produce the afterglow emission
observed in the X-ray and lower energy bands.

Thanks to the rapid spacecraft response to the BAT triggers, the XRT has been able to observe 
and autonomously localize GRB afterglows at early times, 
unveiling new and unexpectedly complex behaviors of the X-ray lightcurves,
which frequently show multiple breaks 
with segments of distinct decay slopes \citep{nousek} and
flares superimposed on the fading afterglow.  

In this paper, we report on the properties of the X-ray afterglow of GRB 050607, 
which was dominated, in its early phase, by emission from X-ray flares superimposed
on the overall decay.  
This was the fourth case in which X-ray flares were detected  in 
the afterglows of GRBs discovered by {\it Swift} 
\citep{patrizia,olivi,abe}.
X-ray flares have been previously observed in few cases by BeppoSAX (XRF 011030,
GRB 011121 and GRB 011211) and were 
identified as the onset of the afterglow \citep{galli,piro}. 
A more detailed study of the characteristics and the
origin of X-ray flares is now possible thanks to the
unique {\it Swift} capabilities, which have greatly increased
the number of rapid observations of the early phase of the X-ray afterglows.
Continued internal engine activity at late times 
was proposed to explain the very short timescale and spectral variations during the flares
observed in XRF 050406 (the first flare detected by {\it Swift}, see \citealt{patrizia}) 
and the very bright flare of GRB 050502b \citep{abe}, which peaked 740 s after the 
GRB prompt emission.
The flares in the X-ray afterglow of GRB 050607 had asymmetric shapes, 
with a steeper rise than previously observed flares, and a shallower decay,
similar to the FREDs (Fast Rise, Exponential Decay) that are frequently observed
in the GRBs prompt emission.

In this work, we describe the XRT data analysis 
of the GRB afterglow, focusing in particular on the temporal and spectral properties
of the early phase of flaring activity of the X-ray emission.
We use the notation $F(t,\nu)\propto (t-t_0)^{-\alpha}\nu^{-\beta}$ 
for the temporal and spectral dependence of the flux.  
All uncertainties are given at the 90\% confidence level for one interesting parameter ($\Delta\chi^2$=2.7).
The observations are presented in \S~\ref{obse} and 
the data analysis in \S~\ref{anal}.
The discussion and conclusion are in \S~\ref{disc}.
A detailed presentation of the procedure applied to correct for the \lq pile-up\rq \
effect, caused by the high X-ray afterglow count rate during the flares is given in the appendix.  

\section{Observations}
\label{obse}

The Burst Alert Telescope triggered on the GRB 050607 
at 09:11:22.81 UT on 2005 June 07 \citep{GCN 3525}.  
The spacecraft started the slew to the GRB 16.5~s after the trigger,
while the burst was still ongoing, 
and settled on the burst 84~s after the trigger.

The XRT follows an autonomous sequence of observations in response
to a GRB trigger, starting with the acquisition of 
a short image to localize the GRB afterglow and then  
performing observations in different readout modes
according to the source intensity \citep{modes}.
Due to engineering tests on the narrow field instruments 
at the time of the trigger, the XRT on-board
centroiding algorithm could not converge on a source when the first
image was acquired and the 
nominal observing sequence could not be effectively followed.
The instrument collected a single frame in Windowed Timing (WT) mode
and then immediately switched into Photon Counting (PC) mode, 
designed to provide two dimensional spatial information at the expense of a 
lower timing resolution.
The GRB afterglow was monitored by the XRT for 30 days 
after the burst for a total exposure time of 363~ks. 
The complete set of the GRB observations is reported in Table~\ref{T1}.

The UVOT was in an engineering mode at the time 
of the GRB event and therefore immediate data were not available.  
The first UVOT observation of the GRB field started
at 22:36:23 UT, 12 hours and 25 minutes after the BAT trigger.  
No optical counterpart was detected \citep{GCN 3537}.

The GRB 050607 field was observed with ground-based telescopes 
in the optical and the IR bands.  
A fading source was detected with the 4m Mayall Telescope at 
Kitt Peak Observatory in the $I$~band with an initial 
magnitude of $21.46\pm0.3$, 640~s after the burst trigger \citep{GCN 3527}.
The transient was observed to have
faded by 2.3$\pm$0.2 mag in the $I$~band  between 640~s and 93.2~ks \citep{GCN 3540}.
Observations were also performed in the $B$, $R$, $z$ and $g$ bands. 
The afterglow was detected in the $R$~band, 
but not clearly visible in the $B$~band images \citep{GCN 3531},  
indicating a possible Lyman break in the spectrum.

The MDM 2.4m telescope also reported a source 
detected in the $R$~band (22.5$<R<$23 mag, \citealt{GCN 3530}) 
with the observation beginning 
16 min after the GRB trigger.  The object appeared 
to have faded (upper limit of $R>$24 mag) 
in a follow-up observation performed on 2005 
June 8 at 08:15 UT, 23 hours after the trigger \citep{GCN 3541}.

Observations were also performed in the IR by the 
REM robotic telescope \citep{GCN 3538}; 
prompt observations (192~s after the trigger) did not 
reveal any new bright object in the $H$~band to a 
limiting magnitude of 16.0.

\section{Data Analysis} 
\label{anal}

\subsection{Gamma-ray Analysis}

The BAT data (analyzed with BAT HEAsoft software v6.0.2)
showed a 3-peak lightcurve with a $T_{90}$ of 26.5~s, close to the mean value for 
long bursts of the BATSE catalog \citep{batse}.
The measured BAT flux had a peak of $1.15\pm0.16$ photon cm$^{-2}$ s$^{-1}$ and 
a total fluence (pre-slew and slew emission) of $5.8\pm0.5\times10^{-7}$~ergs cm$^{-2}$ in 
the 15--150~keV band.  

The BAT spectrum of GRB 050607 is well fitted with a single power law model with 
index $1.83\pm0.14$.
A fit of the spectrum with a Band function \citep{band}~\footnote {$f(E) = K_{30} 
(E/30)^{\alpha} \exp (-E(2+\alpha)/\ep)$,  
if $E < (\alpha - \beta) \ep / (2 + \alpha)$ and 
$f(E) = K_{30}\{(\alpha - \beta)\ep/[30(2+\alpha)]\}^{\alpha-\beta} 
\exp (\beta - \alpha) (E/30)^{\beta}$, if $E \geq (\alpha - \beta) \ep / (2+\alpha)$.}
does not provide a significant improvement in the $\chi^2_{red}$ because of
the difficulties in constraining $\ep$, due to the narrow BAT energy range and
the relative low fluence of the GRB 050607 prompt emission.

The extrapolation of the BAT lightcurve into the XRT energy band 
is performed using the best fit parameters obtained 
from the prompt emission spectral fit.
If one assumes the GRB prompt emission to have 
a \lq universal\rq \ spectral shape \citep{promptpar} 
described by a Band function with $\alpha$=1.0 and $\beta$=2.5,
the photon index of 1.83 of the simple power law model fit of the 
BAT spectrum is very steep and would yield too high a value of the flux
in the XRT energy band.
The prompt flux in the XRT band can therefore be assumed to lie between the fluxes 
extrapolated from the two spectral models: a simple power-law
model and a Band function with fixed $\alpha$=1.0 and $\beta$=2.5.

The BAT spectrum was analyzed separately over different time intervals
relative to the three peaks of the 
lightcurve prompt emission.
In the time interval $T_0$+30~s to $T_0$+100~s there is no source detection
in the BAT observation and a 3~$\sigma$ upper limit was calculated using the 
two spectral models.  The BAT extrapolated fluxes are reported in Table~\ref{T2}.

\subsection{GRB position}

\begin{figure}[ht]
\figurenum{1}
\includegraphics[angle=270,width=0.45\textwidth]{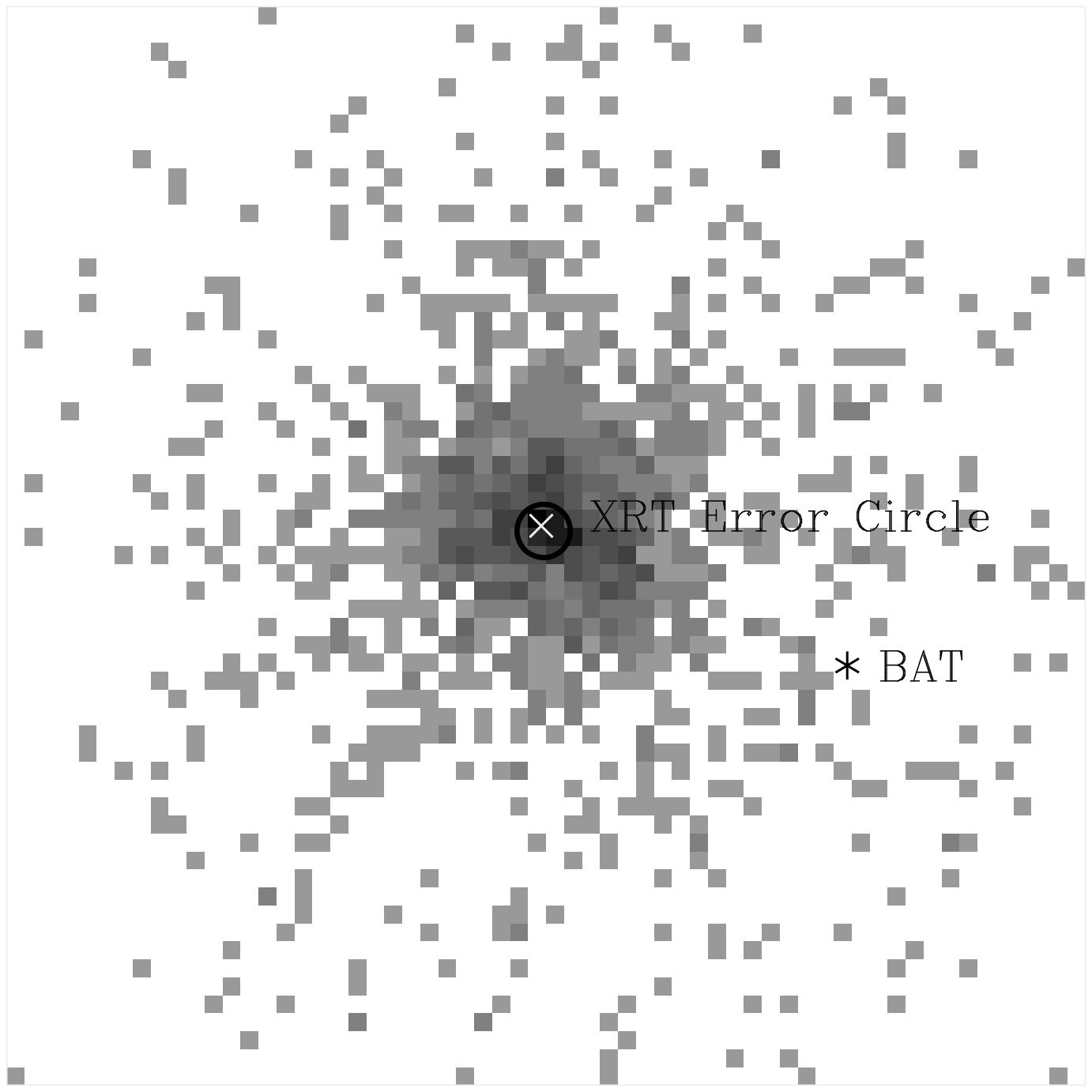}
\caption{The GRB 050607 afterglow XRT image extracted from the first orbit of Photon Counting 
observation.  The BAT refined ground position, the XRT 3.8$\arcsec$ error circle and 
the optical afterglow position (white cross inside the XRT error circle) 
determined with the Mayall Telescope at Kitt Peak \citep{GCN 3527} are shown.}
\label{ima}
\end{figure}

The X-ray afterglow position, 
determined from ground-processed data, 
is $\xrtposra$, $\xrtposdec$, with an uncertainty of
3.8$\arcsec$.
The position obtained with the {\scshape xrtcentroid}
task was corrected for the systematic offsets observed in 
XRT and optical counterparts, caused by the slight misalignment 
between the XRT optical axis and the spacecraft star tracker boresight \citep{corposo}.
The X-ray afterglow is 44$\arcsec$ from the reported BAT position \citep{GCN 3525} 
and 0.9$\arcsec$ from the
optical counterpart reported by \citet{GCN 3527}.
The image of the afterglow detected by the XRT during the first orbit
is presented in Figure~\ref{ima}, along with the BAT refined position \citep{GCN 3525}, the XRT 
error circle and the optical counterpart position \citep{GCN 3527}.  

\subsection{Timing Analysis}
\label{Time}

\begin{figure}[ht]
\figurenum{2}
\epsscale{1.0}
\plotone{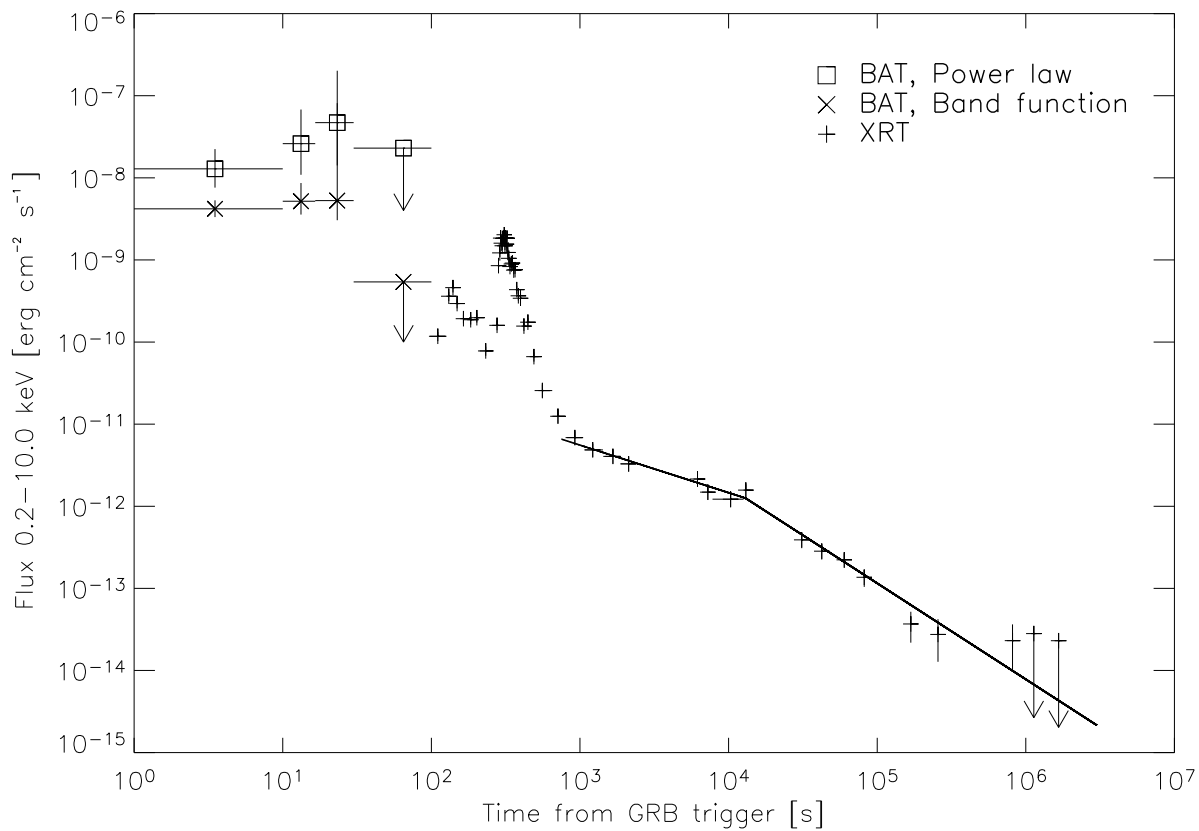}
\caption{The combined BAT and XRT GRB 050607 lightcurve.  The BAT fluxes were extrapolated to the 0.2--10.0~keV 
energy band assuming a simple power law (squares) and a Band (asterisks) spectral model.
In the last BAT time bin (30--100~s) there was no detection in the BAT data and upper limits were derived.
The X-ray afterglow from the entire set of observations is in the 
energy band 0.2--10~keV, pile-up corrected.
The time bars correspond to the time bin sizes.  
The best-fit broken power law model is shown.} 
\label{lc_all}
\end{figure}

The X-ray afterglow of GRB 050607 had a complex behavior, dominated by flares
in the early observations, followed by a phase of slow decay that lasted approximately
12~ks and a late steepening to values typical of GRB afterglows.
The X-ray lightcurve has been extracted 
in the energy range 0.2--10.0~keV 
and binned to obtain 20 background-subtracted counts per bin.
Count rate to flux conversion factors were obtained
from the best spectral model fit for the different afterglow phases
(the detailed spectral analysis is discussed in \S~\ref{bandratio}).
The X-ray lightcurve of GRB 050607 and the
extrapolated BAT fluxes in the 0.2--10.0~keV band are shown in  Figure~\ref{lc_all}.

Due to the high count rate during the flares,
the early XRT observations were severely
affected by \lq pile-up\rq, the result of two 
or more photons being collected 
in a single CCD cell in a readout frame.
In Appendix~\ref{pileup} we describe how the
observations are altered by the  
pile-up effect and we determine the
affected region and the coefficient
applied to the observed fluxes 
to correct for the pile-up. 

\begin{figure}[ht]
\figurenum{3}
\epsscale{1.0}
\plotone{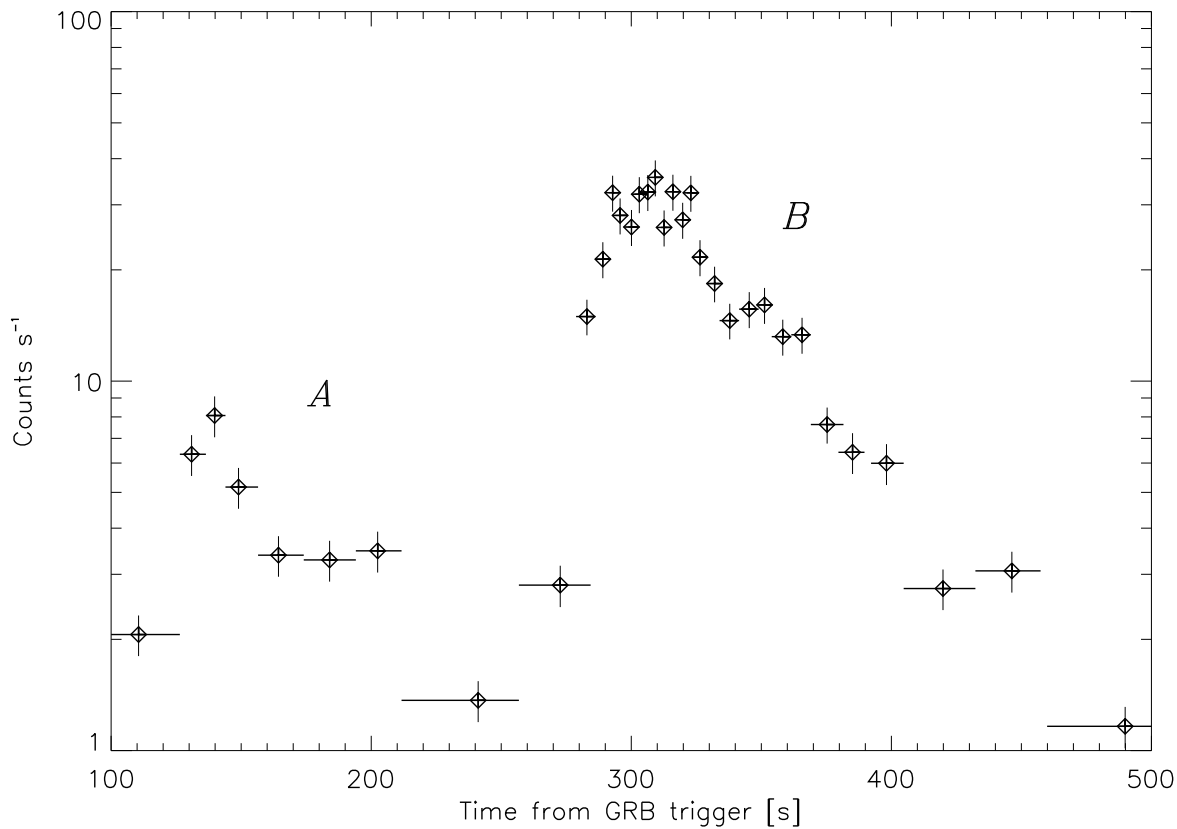}
\caption{The XRT lightcurve of the early GRB 050607 afterglow, in the energy band 0.2--10~kev, 
corrected for pileup, with 20 photons per bin. 
The early X-ray emission was characterized
by the presence of flares.  Flare~A had a peak at $\sim$~8 counts $s^{-1}$,  
the brighter flare~B lasted from 255 to 510~s 
after the first trigger with emission around 30 counts $s^{-1}$ for
approximately 30~s. 
Both flares are asymmetric:  the decay is shallower than the
very steep rise.  This may be due to additional short timescale activity 
superimposed on the main flare, like the small \lq bump\rq \ observed at approximately 355~s 
after the burst trigger.}  
\label{lc_orbit1}
\end{figure}

The phase of flaring activity of the GRB 050607 X-ray afterglow 
is shown in Fig.~\ref{lc_orbit1}.
To avoid the region contaminated by
pile-up, the lightcurve of the 
first 510~s of observations after the GRB trigger
was extracted from an annulus
that excluded events in the central 
8~pixels (18.88$\arcsec$) and  with an outer radius of 40~pixels.
The first flare (referred to as flare \lq A\rq) was observed from 96~s to 255~s after the GRB trigger, 
peaking at
4.6$\times 10^{-10}$~erg ${\rm cm}^{-2} {\rm s}^{-1}$ (8.1~counts~${\rm s}^{-1}$) 
approximately 135~s after the trigger.  
The second flare (flare \lq B\rq) was the 
brightest, lasting from 255~s to approximately 510~s after the trigger and 
including the majority of the X-ray emission, with a peak flux 
of 2.0$\times 10^{-9}$~erg ${\rm cm}^{-2} {\rm s}^{-1}$ (35.6~counts~${\rm s}^{-1}$) $\sim$ 310~s 
after the trigger and a total fluence of
1.36$\times 10^{-7}$~erg ${\rm cm}^{-2}$ ($\simeq$23\% of the BAT prompt emission). 

The flares were asymmetric,
similar to the FREDs that are often observed in the
prompt GRB emission, with a rising side steeper than the following decay.
In the case of flare~A, the peak was reached in $\sim$~30~s, with the 
decay taking approximately 130~s, and for flare~B the rising phase lasted for 
$\sim$~35~s, followed by a \lq plateau\rq \ at the flare maximum with the X-ray emission 
at a countrate of $\sim$~30 counts ${\rm s}^{-1}$ for approximately 30~s,
and a decaying phase $\sim$190~s long that presents 
marginal indications of
spikes and \lq bumps\rq.

The flare's rise and decay can be fitted with a 
simple power law.
With the notation $(t-t_0)^{\alpha}$, 
fixing $t_0$ at the onset of the flares,
the flare~A slope indices are $\alpha_{A,rise}$=1.2$\pm$0.2
and $\alpha_{A,decay}$=-1.3$\pm$0.3, and for flare~B are 
$\alpha_{B,rise}$=4.1$\pm$0.8
and $\alpha_{B,decay}$=-2.3$\pm$0.3.
The choice of $t_0$ is critical in 
determining the slopes indices. In fact, if $t_0$ is 
fixed at the GRB trigger time $t_B$ the slopes are steeper, yielding
higher values than for previously detected flares \citep{patrizia,abe}.
In this latter case flare~A rises as $(t-t_B)^{5.9\pm0.6}$ and decays
as $(t-t_B)^{-2.8\pm0.6}$, while for flare~B the rise and decay are
fitted with $(t-t_B)^{22\pm5}$ and  $(t-t_B)^{-7.2\pm0.5}$.

The brighter flare~B decay was fitted excluding the data points
of the secondary bump superimposed on the flare and 
centered 355~s after the burst trigger.
From the power law fit decay parameters, 
the expected number of events in that time period is 323$\pm$18,
while the observed number of events is 398$\pm$20.
The observed 75 excess photons are significant at 
a 2.8 $\sigma$ level.
Although hints of short time scale structure in the XRT lightcurve are present, 
in particular during the brightest flare, no significant periodicity was found 
from the timing analysis in the $2-300$~s period
range, using a frequency spacing of $10^{-3}$~s$^{-1}$.

After the flares, the X-ray afterglow emission decayed to intensities for which the 
effect of pile-up was negligible.  Data were extracted 
using a circle of 20 pixels radius (smaller than the one
used for the flares to avoid excessive background contamination)  
centered on the GRB afterglow and the background was evaluated using a 
circle with $r$=40~pixels in an area of the CCD without presence
of other sources.
In the second, third and fourth observations 
(segments 1, 2 and 3 respectively, see Table~\ref{T1}) a binning of 25 events per bin was used.   
At later times multiple observations (segments) were co-added to detect the X-ray afterglow,
and a circle of 4~pixels radius centered on the GRB position was used to obtain a 
higher signal-to-noise ratio.  The measured afterglow flux was PSF-corrected to account
for the smaller extraction region.
The last two data points of the lightcurve are 90\% upper limits (Fig.~\ref{lc_all}).

After the main flare, starting from 510~s after the GRB trigger, the afterglow is best fitted with a broken
power law (shown in Fig.~~\ref{lc_all}) with indices of $0.58\pm0.07$ and 
$1.17\pm0.07$, with a break time of 11.7$\pm$0.6~ks ($\chi^2_{red}=1.02$).  
The broken power law model is a significantly better fit (F-test probability of 4.8x$10^{-3}$)
to the late X-ray afterglow than a single power law (index of $0.88\pm0.03$ with a $\chi^2_{red}$ of 1.91).  
The observed decay slopes and the degree of steepening after the break are close to the average values
obtained from the sample of 27 GRBs discovered by {\it Swift} reported in \citet{nousek}.

\subsection{Spectral Analysis}
\label{bandratio}

The X-ray afterglow of GRB 050607 can be divided into four time intervals:  
flare~A extending to 255~s after the trigger, the stronger
flare~B from 255~s to approximately 510~s, the shallow section 
of the afterglow emission and the 
late steepening after 12~ks.  
The spectral analysis (using  {\scshape xspec} v11.3.2) was performed 
separately on each lightcurve phase, selecting events with grades 0 to 12.
Because of low statistics below 0.5~keV and at high 
energies the spectra were fitted in the
energy range 0.5--6.0~keV;  
the swxpc0to12\_20010101v007.rmf response matrix file 
from the {\it Swift} 
XRT calibration database was used for the spectral fit.
The ancillary file was created using the {\scshape xrtmkarf} command 
of the {\scshape xrtpipeline} software, applying the PSF correction. 
The spectra during the first 510~s of observation 
were extracted from an annular region
as in the timing analysis
to avoid pile-up contamination.   
The background was evaluated from a circle of $r$=40~pixels
in a region without sources.

Due to the low number of events during the first flare 
and the late part of the lightcurve, during these time intervals
C-statistics \citep{cash} 
were used in the fitting procedure, while $\chi^2$ statistics were used for the main flare.  
Since the X-ray count rate at low energy was too low to properly constrain the 
X-ray absorption column, the $N_H$ was fixed to 1.41$\times 10^{21}~{\rm cm}^{-2}$, 
the weighted average value 
in the GRB 050607 direction \citep{nh}.   
The flares were fitted with both a simple absorbed power law and  
a cutoff power law model to investigate a possible common origin of the
X-ray flares and the prompt gamma-ray emission: in the energy band 
of the XRT data a cutoff power law is almost equivalent to a Band function.

The best fit of flare~A with an absorbed power law model 
yielded a photon index of $2.00^{+0.19}_{-0.18}$, 
with the absorption fixed at the Galactic value. 
The brighter flare~B fit with an absorbed power law model yielded an absorption 
column density $2\sigma$ higher than the Galactic value.
To allow comparisons with earlier and later spectra, the absorption was 
fixed to the Galactic value,
obtaining a photon index of $2.27^{+0.13}_{-0.12}$, while selecting
only the decaying phase of the flare yielded a photon index
of $2.31^{+0.21}_{-0.19}$.

\begin{figure}[ht]
\figurenum{4}
\epsscale{1.0}
\plotone{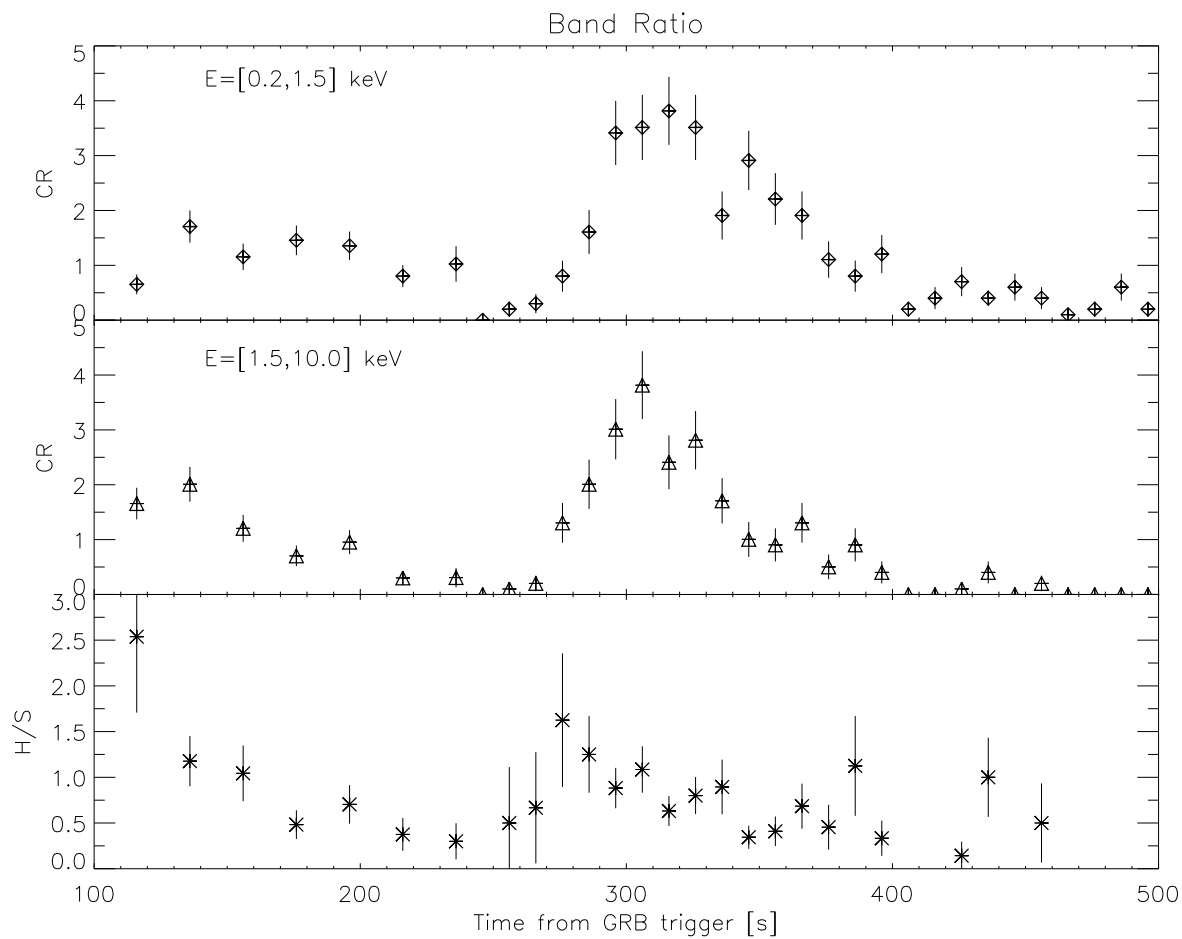}
\caption{The band ratio of the early phase of the X-ray afterglow.  
The hard (H=1.5--10.0~keV) and the soft (S=0.2--1.5~keV) band energy ranges were chosen
to obtain a similar number of counts in the two bands.  The band ratio
(H/S) shows evidence of spectral evolution during the two flares, with a harder
emission at the flare's onset and a soft component that lasted longer and dominated 
the decay of the flare.}
\label{hratio}
\end{figure}

The fit of flare~B with a cutoff power law model yielded 
a best fit absorption column density
closer to the Galactic value than in the single power law case.
Fixing the absorption to the Galactic value, the cutoff power law model is a better fit
than an absorbed power law ($\chi^2_{red}$ improved from 1.54 to 1.14).
The parameters of the cutoff power law model fit are not very well constrained, especially in the case
of the cutoff energy, due to the limited 
number of photons in the dataset at high energies.
The cutoff power law model fit yielded a photon index of $1.41^{+0.48}_{-0.53}$,
harder than the simple power law, and a cutoff energy of $2.4^{+3.5}_{-0.9}$~keV.
Our attempt to also fit the dimmer, first flare with a cutoff power law did
not properly constrain the fit parameters due to the low number of photons during
that time interval. However, 
we investigated the 
spectral evolution of the afterglow during the flares, dividing 
the lightcurve into a low energy band and a high energy band and evaluating the 
band ratio.  
The energy ranges (0.2--1.5 and 1.5--10.0~keV for the low
and the high energy bands respectively) were selected to obtain a similar number of counts in the two bands.  
Spectral evolution is evident in the band ratio plot (see Figure~\ref{hratio}). 
In particular, 
the hard band contributed to the overall emission mostly during the rising
part of the two flares, while the soft emission lingered longer and started
decaying at later times.  The observed behavior 
indicates that the emission during the flares is harder than the underlying 
afterglow, and that the flares evolve spectrally, softening as they decay.

\begin{figure}[ht]
\figurenum{5}
\epsscale{0.8}
\includegraphics[angle=270,width=0.85\textwidth]{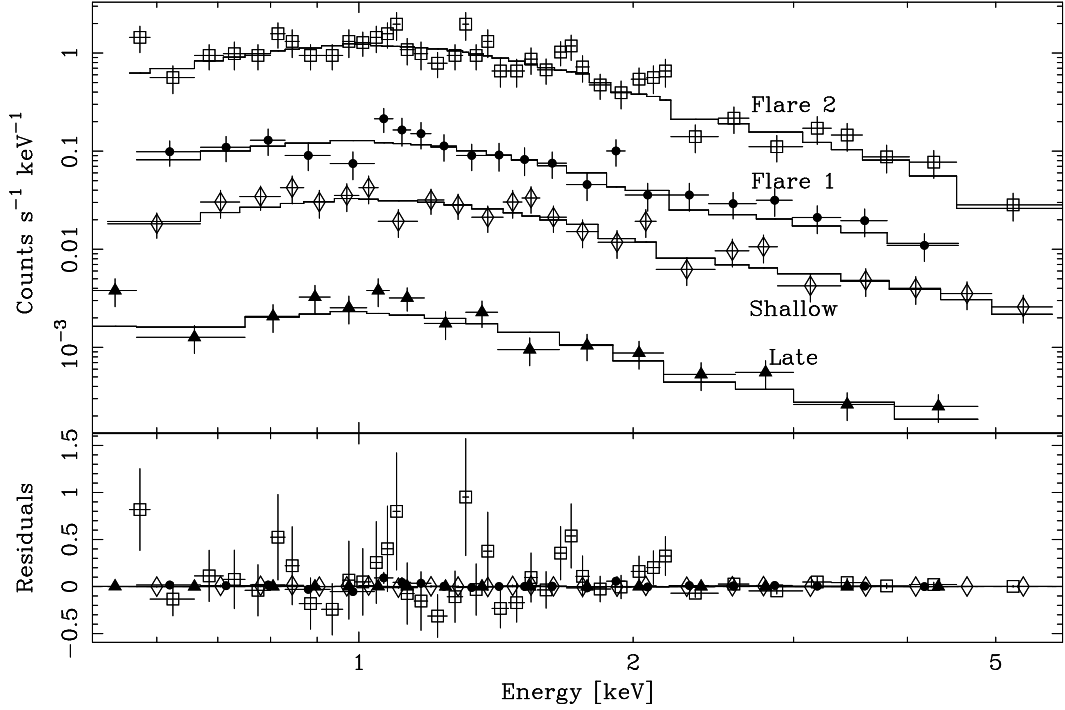}
\caption{The XRT spectra of the different afterglow phases.
The data were fitted in the energy range 0.5--6.0~keV with a binning of 20~photon per bin
for the bright second flare and a binning of 1~photon per bin for the other phases,
for which C-statistics was used.  The data are rebinned here to
10~photons per bin
for presentation purposes, and the 
first flare spectrum is normalized by a constant factor of 0.2 for clarity. 
The absorbed cutoff power law model was used to fit the spectrum of the bright flare while 
a simple absorbed power law model was used for the other afterglow phases.
The cutoff power law model yielded a harder spectrum ($\Gamma$=1.41) for the bright flare 
than the late afterglow.  
In the lower panel the differences between the data and the fitted models are displayed.}
\label{spec2}
\end{figure}

The late part of the X-ray afterglow was analyzed using 
events from a circular region of 20 pixels radius.
We applied C-statistics to
fit the data with an absorbed power law with absorption fixed to the Galactic value,
yielding a photon index of $1.78^{+0.18}_{-0.13}$ for the shallow section
of the lightcurve
and a photon index of $1.97^{+0.36}_{-0.34}$ after the lightcurve break.   
The spectra of the different afterglow phases are shown in Figure~\ref{spec2} and the
fit results are presented in Table~\ref{T3}. 

\section{Discussion and Conclusions}
\label{disc}

The X-ray lightcurve of GRB 050607 has a complex behavior,
presenting most of the unexpected features 
that have been observed in the afterglows of GRBs
discovered by  {\it Swift}, such as the presence of multiple breaks
in the lightcurve and the flaring activity superimposed on the 
overall lightcurve decay.

XRT observations indicate that flares are common 
in the early phase of X-ray afterglows 
\citep{Burrows,X-ray-theo,nousek}:
as of 2005 October 20, flares were detected 
by the XRT in 23 of the 50 afterglows for which observation
with the narrow field instruments started within 10
minutes from the BAT trigger.
Several scenarios have been proposed as the possible source of
flux variabilities of the X-ray afterglows:  
density fluctuations in the enviroment \citep{lazzati}; angular structures in the outflow \citep{nakar}; 
refreshed decelerated shocks by slower shells \citep{rees}; synchrotron emission
from a reverse shock \citep{shiho}.
The time variability for these models, all related to external shock emission,
is longer than the observed time since the GRB trigger \citep{ioka}. 
Therefore, they are unable to account for the very 
short timescales of the two flares of GRB 050607, that showed in particular a very
steep rise, sharper than for previously 
reported cases \citep{patrizia,abe}, with values of $\delta t_{rise}/t_{peak}\leq0.2$,
and slower dacays with $\delta t_{decay}/t_{peak}\leq$~1. 
The flares of GRB 050607, unlike the ones previously 
detected by the XRT, did not have a symmetric shape but were rather 
similar to the FREDs that are common
in the prompt GRB emission, and had indication of additional 
short time variability superimposed on the flares emission.

The choice of the spectral model to fit the X-ray flares observed is  
important for a correct interpetation of the data: in the case of GRB 050607
the cutoff power law model (coincident to a Band function in the XRT energy band)
provided a significantly better fit of 
the brighter flare spectrum than a simple absorbed power law.
In addition, the hardness ratio showed that the emission was harder 
at the flare onset, while the decay was characterized by a softer component.
In the late flares detected by BeppoSAX, the flare spectral properties were
similar to those of the late afterglow, and they were therefore interpreted as 
the onset of the afterglow itself. 
The spectral evolution observed during the flares of GRB 050607 
suggests instead a mechanism for the origin of the flares 
distinct from the one responsible for the emission of the underlying decaying afterglow.

The described characteristics observed in the flares of GRB 050607 strongly indicate 
internal shock emission from late central engine activity
at a few hundred seconds after the prompt gamma-ray burst as the origin of the flares. 
Compared with the three peaks in the prompt emission detected by the BAT, 
the X-ray flares had lower energy and longer time scale.
If the late internal shocks happened
at larger radii than those of the prompt gamma-ray emission, smaller 
magnetic field strength (internal energy density) and a larger 
curvature time scale (angular spreading time scale) for the shocks
would make the resultant flares softer and longer, 
consistent with our observations.
At larger radii the fireball shells should spread and have a larger width,
and the observed plateau of the bright flare could be interpreted as produced 
by a thick shell collision or by a series of collisions.

For a large fraction of {\it Swift} GRBs, the X-ray afterglow is 
characterized by a steep to shallow to \lq normal\rq \ behavior \citep{chinca, taglia,
X-ray-theo,nousek,Panaitescu}. 
In the case of GRB 050607 the flares are presumably superimposed on 
the steep decay and the transition to the shallow phase is not 
well constrained.
A shallow to \lq normal\rq \ transition was observed 
$\sim$~12ks after the burst. 
The steepening ($\alpha$=0.58$\pm$0.07 to 1.17$\pm$0.07) is 
larger than the value of $\delta\alpha$=1/4 expected when the cooling frequency 
passes through the observed band in the standard constant energy blast 
wave model and therefore implies a change of the hydrodynamics of the forward
shock, suggesting that the forward shock continues to be refreshed for 
some time \citep{X-ray-theo,nousek}. 
In one possible mechanism, the bulk
Lorentz factor of the ejecta decreases at the end of the prompt phase, resulting in a monotonic increase
of the Lorentz factor with radius.  This outflow gradually catches up
with the forward shock (blast wave), and gives a smooth energy 
injection \citep{Sari,Panaitescu}. In another scenario, 
the energy injection originates from 
a long-lived central engine, active up to a few hours after the
prompt emission \citep{dai}.
The decay slope after the break ($\alpha$=~1.17) is consistent with the
standard interstellar medium (ISM) 
forward shock model \citep{sariwind}. If we apply the standard blast wave model to explain the 
light curve after 12ks, an electron index p~$\sim$~2.2 is obtained from the decay 
and spectral indices.

If one assumes that the shallow decay observed in the X-ray energy band originates from
refreshed shocks, emission from the associated reverse shocks is expected
to peak in the optical band \citep{Sari}, with a decay slightly slower than the forward shock emission.  
The $I$ band decay during the first 6400~s 
could be modeled with a single power law of index 0.5  \citep{GCN 3531}.
The measured $I$ magnitudes and the X-ray fluxes are 
consistent within the uncertainties with the model predictions during the shallow phase.
With the late measurement of the transient, still detectable in the $I$ band
93~ks after the burst trigger \citep{GCN 3540},
the $I$~band afterglow decay could be modeled with a single power law decay with slope 0.4, or 
0.3 if the first detection after 640~s was excluded.
The fit of the X-ray afterglow decay after the flares with a single power law  
yields a slope of 0.88.  The difference between the reported $I$~band decay and the 
X-ray slope suggests that the late $I$~band detection had a significant
contribution coming from the GRB host galaxy.  
A new deep observation of the field with ground based telescopes would
verify this interpretation.

\acknowledgments
This work is supported at Penn State by NASA contract NAS5-00136; 
at OAB by funding from ASI on grant number I/R/039/04;
and at the University of Leicester by the Particle Physics and Astronomy
Research Council (PPARC).  
We gratefully acknowledge the contributions of dozens of members of the {\it Swift} team at PSU, 
OAB, University of Leicester, GSFC, ASDC and our subcontractors, 
who helped make this Observatory possible, and to the 
Flight Operations Team for their support above and beyond the call of duty.  

\appendix

\section{Pile-up correction}
\label{pileup}

The pile-up effect 
causes two or more photons to be considered by the CCD as a single event 
with energy equal to the sum of the energies 
of the individual photons.
Pile-up, therefore, has a major impact on the observations, by lowering the apparent source count rate 
and making the apparent spectral index harder 
(for a detailed discussion see for example \citealt{Weisskopf}).  

Pile-up  also modifies the grade distribution of the photons collected on the CCD, 
via an effect known as \lq grade migration\rq \ \citep{Davis}.  
When an X-ray photon hits the CCD surface, the charge produced is spread over one or more pixels; 
single pixel events are recorded as \lq grade 0\rq \ events, 
split pixel events are designated with higher grades (1 to 12 for X-ray events, 
see \citealt{Burrows_xrt} for the XRT grade definitions).
The effect of pile-up is to decrease the number of single pixel events and 
to increase the number of apparent multiple pixel events, depending on 
the overall incident count rate.

To investigate the extent of the CCD region affected by pile-up
we derived the grade distribution of the events collected from the GRB afterglow during the main flare,
when the XRT count rate exceeded 1~count ${\rm s}^{-1}$.  
Annular regions centered on the GRB emission with increasing 
inner radii and fixed outer radii ($r_{out} = 40$~pixels) were extracted to verify the 
spatial extent to which the observations are affected by pile-up. 
The ratio of single pixel events to the total number of events 
has been calculated for each annular region  
and the results are reported in Table~\ref{T4}. 
For comparison, ground calibration tests \citep{Moretti} with the XRT showed 
that in non piled-up regimes the fraction of single pixel events is 0.78 at 1.49~keV 
(the mean photon energy during the bright flare, once the pile-up correction 
was performed, was measured to be 1.7~keV).

For annular regions with inner radii less than 5~pixels the \lq grade migration\rq \ effect 
caused by pile-up is significant, 
while for inner radii larger than 8 pixels the fraction of single pixel events 
remained constant and became comparable with ground calibration
measurements for sources with low count rates.  
To confirm the result, the radial profile of the bright flare was extracted and compared with
the analytical PSF described in \citealt{Moretti}.  The two profiles showed little or no deviations 
at radii larger than 8 pixels.   
To avoid the region affected by pile-up while minimizing  
the number of rejected events in the central region, 
the analysis of the GRB afterglow during the
bright flare was performed using an annular region that 
excluded events in the central 8~pixels (18.88$\arcsec$) 
and with an outer radius of 40~pixels.

The number of events rejected from the central 8 pixels is in principle 
dependent on the incident spectrum and the details of the energy-dependent PSF.
For a point source at 2$\arcmin$ from the center of the CCD (as 
is the case of the GRB 050607 X-~ray afterglow during the first orbit), 80\% of the PSF 
is included in an 8 pixel circle at an energy of 0.5~keV, 77\% at 1.7~keV (the mean energy 
of the flare spectrum corrected for pile-up) and 74\% at 3.0~kev.
We therefore used {\scshape xspec v11.3.2} to calculate the correction factor for the pile-up.
The spectra from the annular region and the background region were extracted;
the {\scshape xrtmkarf} command (version 0.4.14) 
was used to create two ancillary response files, 
with and without applying the PSF correction.
Fitting the spectrum with an absorbed power-law model we obtained 
the best-fit parameters for the two ancillary response files. The ratio 
of the two normalization parameters yields 
the correction factor to 
account for the effect of pile-up . 
The coefficient ($c_{pileup}=0.20\pm0.03$) obtained from the spectral analysis was used to
multiply the count rate observed in the annular region and
is consistent  with the fraction of 
PSF contained in the 8 pixel central region.

\clearpage

\clearpage

\clearpage

\clearpage

\clearpage 

\clearpage

\begin{deluxetable}{cccccc}
\tablecaption{Observations of GRB 050607
\label{T1}}
\tabletypesize{\normalsize}
\tablecolumns{5}
\tablewidth{0pt}
\tablehead{
\colhead{Sequence Number}  &
\colhead{Number of orbits}  &
\colhead{Start Time} &
\colhead{Stop Time} &
\colhead{Exposure Time}\\
\colhead{ }&
\colhead{ }&
\colhead{UT}&
\colhead{UT}&
\colhead{[s]}
}
\startdata
0  & 3  &  2005-158-09:12:56   & 2005-158-13:04:40  &  7955    \\
1  & 13 &  2005-158-17:13:52   & 2005-159-13:11:19  & 44423    \\
2  & 11 &  2005-160-00:05:00   & 2005-160-16:30:51  & 30730    \\
3  & 13 &  2005-161-00:05:00   & 2005-161-23:03:17  & 26397    \\
4  &  8 &  2005-162-00:04:59   & 2005-162-21:03:57  &  4556    \\
5  &  3 &  2005-165-19:38:15   & 2005-165-23:29:47  &  7721    \\
6  & 14 &  2005-167-00:36:29   & 2005-167-23:42:54  & 31998    \\
7  & 12 &  2005-168-02:17:51   & 2005-168-23:44:58  & 24316    \\
8  & 29 &  2005-169-23:17:40   & 2005-171-23:59:57  & 39943    \\
9  & 14 &  2005-172-01:22:51   & 2005-172-23:59:59  & 12775    \\
10 & 39 &  2005-176-01:34:52   & 2005-178-21:41:30  & 62773    \\
11 & 12 &  2005-181-01:01:56   & 2005-181-23:37:51  & 12189    \\
12 &  3 &  2005-182-01:07:13   & 2005-182-12:29:26  &  4160    \\
13 &  3 &  2005-182-17:15:46   & 2005-182-20:31:42  &  1682    \\
14 & 11 &  2005-182-23:41:26   & 2005-183-23:51:23  &  7676    \\
15 &  9 &  2005-184-01:23:59   & 2005-184-22:21:44  &  6979    \\
16 & 21 &  2005-184-23:45:48   & 2005-186-12:55:57  & 21884    \\
17 &  6 &  2005-186-14:25:03   & 2005-186-22:35:26  &  5005    \\
18 & 11 &  2005-187-00:04:59   & 2005-187-22:42:18  & 10566    \\ 
\enddata
\end{deluxetable}

\clearpage

\begin{deluxetable}{ccccc}
\tablecaption{GRB 050607. BAT extrapolated Fluxes 
\label{T2}}
\tabletypesize{\normalsize}
\tablecolumns{5}
\tablewidth{0pt}
\tablehead{
\colhead{Time range}  &
\colhead{Power-law Flux} &
\colhead{$\chi^2_{red}$} &
\colhead{Band Function Flux} &
\colhead{$\chi^2_{red}$}
\\
\colhead{[s]}           &
\colhead{[$10^{-10}$ erg ${\rm cm}^{-2} {\rm s}^{-1}$]} &
\colhead{} &
\colhead{[$10^{-10}$ erg ${\rm cm}^{-2} {\rm s}^{-1}$]}&
\colhead{} 
}
\startdata
-3.0--10.0  & $128.75^{+94.96}_{-52.68}$   &0.89 & $41.95^{+11.27}_{-8.70}$   &  0.93 \\ 
10.0--16.5  & $260.58^{+41.85}_{-15.16} $  &1.21 & $51.89^{+34.50}_{-16.19}$  &  1.27   \\
16.5--30.0  & $470.14^{+154.64}_{-328.53}$ &0.73 & $52.80^{+75.91}_{-22.30}$  & 0.72  \\
30.0--100.0 & $<$230                       & & $<$ 0.54                       &   \\ 
\enddata
\tablecomments{GRB 050607: extrapolation into the 0.2--10.0~keV band of 
the BAT data from the prompt emission; two spectral models were assumed:  
a power-law and a Band
Function. The $\chi^2_{red}$ of the spectral model best-fit is reported.
For the fourth time interval, where no source is detected in the BAT data, 
an upper limit was derived for each of the two spectral models.}
\end{deluxetable}
\clearpage

\begin{deluxetable}{lccccccc}
\rotate
\tablecaption{GRB 050607. X-ray afterglow: spectral and timing parameters
\label{T3}}
\tabletypesize{\small}
\tablecolumns{8}
\tablewidth{0pt}
\tablehead{
\colhead{Phase}  &
\colhead{Time from trigger} &
\colhead{$\Gamma$} &
\colhead{$\chi^2_{red}$} &
\colhead{$\Gamma$} &
\colhead{$\chi^2_{red}$} &
\colhead{$\beta$} &
\colhead{$\alpha$ (decay slope)} \\
\colhead{ }&
\colhead{ }&
\colhead{Power Law}&
\colhead{ }&
\colhead{Cutoff Power Law}&
\colhead{ }&
\colhead{Power Law}&
\colhead{ } 
}
\startdata
First flare        & 96s$<$T$<$255s & $2.00^{+0.19}_{-0.18}$ & 142.2\tablenotemark{a}&  \nodata               &\nodata&   \nodata            &   \nodata                        \\
Second flare       & 255s$<$T$<$510s& $2.27^{+0.13}_{-0.12}$ & 1.54                  & $1.41^{+0.48}_{-0.53}$ & 1.14 &    \nodata            &   \nodata                         \\
Second flare decay & 315s$<$T$<$510s& $2.31^{+0.21}_{-0.19}$ & 0.89                  &    \nodata             &\nodata&$1.31^{+0.21}_{-0.19}$ & 7.20 (2.26 if $\rm {T_0}=T_{flare}$)\\
Flat               & 510s$<$T$<$12ks& $1.78^{+0.18}_{-0.13}$ & 150.5\tablenotemark{a}&    \nodata             &\nodata&$0.78^{+0.18}_{-0.13}$ & $0.58^{+0.07}_{-0.07}$           \\
Break              & T$>$12ks       & $1.97^{+0.36}_{-0.34}$ & 127.8\tablenotemark{a}&    \nodata             &\nodata&$0.97^{+0.36}_{-0.34}$ & $1.17^{+0.07}_{-0.07}$           \\
\enddata
\tablenotetext{a}{Value obtained with C-statistics fit}
\tablecomments{Spectral and timing parameters of the X-ray afterglow evolving phases.
The spectra were extracted in the energy range 0.5--6.0~keV with an absorbed power law fit.
The $N_H$ was fixed to the Galactic value of 0.14x$10^{22} {\rm cm}^{-2}$ due to insufficient
statistics at low energies to properly constrain the
X-ray absorbing column.
We investigated the possibility of internal shocks (the mechanism responsible for the prompt emission)
as the origin of the bright second flare, modelling the 
spectrum with an absorbed cutoff power law and obtaining a better fit to the data than using a simple
power law.
As can be seen from the photon indices, the choice of the model is critical for determining the hardness
of the spectrum.  Not enough counts were collected during the first flare to constrain the absorbed cutoff power law
parameters in that time interval, and a simple absorbed power law was used.}
\end{deluxetable}

\clearpage

\begin{deluxetable}{cccc}
\tablecaption{GRB 050607.  Grade distribution\label{T4}}
\tabletypesize{\normalsize}
\tablecolumns{4}
\tablewidth{0pt}
\tablehead{
\colhead{Inner Radius}  &
\colhead{Events} &
\colhead{Single pixel events} &
\colhead{Fraction of total events}
}
\startdata
0 pixel     &   1774  &  1131   &  63.7\%   \\
5 pixels    &   1092  &   805   &  73.7\%   \\
6 pixels    &    958  &   726   &  75.8\%   \\
7 pixels    &    806  &   631   &  78.3\%   \\
8 pixels    &    718  &   569   &  79.2\%   \\
9 pixels    &    634  &   511   &  80.6\%   \\
10 pixels   &    567  &   456   &  80.4\%   \\
15 pixels   &    362  &   292   &  80.6\%   \\
\enddata
\end{deluxetable}
 \clearpage

\end{document}